# Meta-experiments: Improving experimentation through experimentation

Melanie J.I. Mueller (melanie.mueller@booking.com)
Booking.com, Amsterdam, The Netherlands
June 2024

**Abstract**
A/B testing is widely used in the industry to optimize customer facing websites. Many companies employ experimentation specialists to facilitate and improve the process of A/B testing. Here, we present the application of A/B testing to this improvement effort itself, by running experiments on the experimentation process, which we call "meta-experiments". We discuss the challenges of this approach using the example of one of our meta-experiments, which helped experimenters to run more sufficiently powered A/B tests. We also point out the benefits of "dogfooding" for the experimentation specialists when running their own experiments.

## Introduction

Online experimentation is widely used in the industry to learn which product changes are beneficial and should be shipped to customers [1,2]. Companies often run hundreds of experiments at the same time, with experimentation teams taking care of and improving the experimentation process. Running A/B tests should be easy, fast and of high quality.

Experimentation teams can have Key Performance Indicators (KPIs) such as the number of experiments, the satisfaction of experimenters, or the quality of experiments [3,4]. To see whether their actions are successful, experimentation teams generally rely on qualitative feedback from leadership and experimenters, or on timelines of quantitative KPIs such as number of experiments per week. However, these approaches make it difficult to causally attribute improvements to interventions and to quantify impact reliably. We therefore propose that experimentation teams do as product teams do to attribute and quantify impact: run A/B tests. We call these experiments on the experimentation process itself "meta-experiments".

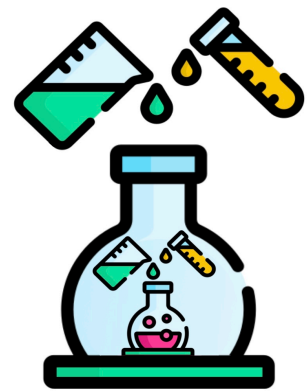

Such meta-experiments do not seem common in the industry; we are not aware of any examples in the literature. This may be caused by the simple habit of not running experiments on "in-house" tools or processes, or no clear metrics to optimize. The main reason might be low sample size - even large companies "only" run hundreds of experiments at a given time [1,2], while website traffic for "normal" A/B tests can be orders of magnitude larger. Here, using the example of one of our meta-experiments which helped experimenters to run sufficiently powered A/B tests, we show how to overcome these challenges, and also discuss pitfalls and potential biases associated with the choice of randomization units and metrics. We use the structure of PICOT (population, intervention / comparison, outcome, time [5]), to present the meta-experiment design, execution, and outcome.

## Intervention / comparison: Example meta-experiment of low-power alerts

Meta-experiments can be run for a variety of interventions. Examples include improving the UX of experiment creation to increase experimenter satisfaction, presenting A/B testing success stories to increase experiment usage, enhancing the results display to improve experiment decision quality, etc.

The main KPI of our experimentation team is experimentation quality [4]. One component of experiment quality is sufficient statistical power of the A/B test. However, some of our experiments are underpowered. Investigation of experimenter behavior, support questions and interviews, showed two main reasons: lack of understanding of the concept of power and its business consequences, and practical difficulty in doing power calculations correctly. We therefore designed an intervention where we pushed an alert to experimenters when their experiment was underpowered at 1 week of runtime. In this alert, we provided an explanation of power and its business impact, as well as a link to set experiment parameters for sufficient power with one click (Figure 1). In the control group, experimenters did not receive such an alert. Here, we will use this meta-experiment as the leading example for discussion.

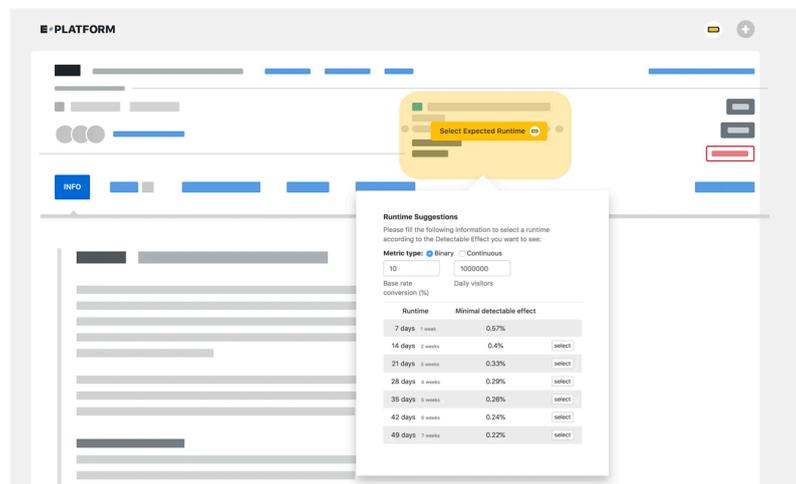

*Figure 1: Schematic representation of the treatment in variant: When their experiment is underpowered, experimenters receive an alert with an easy way to sufficiently power their experiment.*

## Population: Randomization unit and potential biases

When experimenting in a new area, one has to choose the randomization unit and success metrics, which are intertwined choices. For meta-experiments, the most obvious choices for the randomization unit are experiments or experimenters, which both have pros and cons.

Since the actors to be influenced for better experimentation are the experimenters, experimenters are a natural randomization unit. In most cases, experimentation is a team effort and the randomization unit could be the experimenting product team. This unit works well for interventions that target the number of experiments or user satisfaction. Success here can be measured as the number of experiments per team or as a user satisfaction score. However, biases can arise due to variant spillover when different product teams collaborate to run experiments, or when experimenters present experiment results to each other. If the experimentation experience is very different, this can also lead to user confusion, e.g. in the case of vastly different results displays or best-practice guidance.

In our example meta-experiment of power alerts, our success metric is on the experiment level: Is the experiment sufficiently powered or not? Experimenters may run multiple experiments during our meta-experiment; so defining success on experimenter level can be quite difficult. An experimenter may have some underpowered and some not, and some with power 'fixes' and some not. Hence, we chose experiments as the randomization unit, so that our success metric can be sufficient power for each sample experiment. This randomization unit was the main reason to coin the term "meta-experiment" to distinguish our meta-experiment from the sample experiments.

A disadvantage of using experiments as the randomization unit is that they do not correspond to the actors, which are the experimenters. When the same experimenter runs many experiments during the meta-experiment, the resulting variant spillover due to these experiments being distributed randomly over base and variant can lead to user confusion and biased results. In our case, we expect this bias to "amplify" the impact of our treatment, see below.

## Time: Dealing with low sample sizes

We believe that one of the main reasons for the lack of meta-experiments is concern about low sample size and therefore low power. Even large companies "only" run hundreds of experiments at a given time [3], which is small compared to website traffic, which is often orders of magnitude larger. However, smaller companies and startups also run A/B tests with smaller traffic, and medical studies often require less than 100 participants in total, and even for measuring treatment effect seldom more than 200 [5]. It therefore should be possible to run meta-experiments given tens or hundreds of experiments, or similar numbers of experimenters.

In our case, we successfully ran meta-experiments on total sample sizes of 100 to 400 experiments. Before committing, we made sure that power was reasonable by investigating past experiments to see the "traffic" of underpowered experiments, and by doing user research to find interventions with potentially large effect sizes.

Generally, we used the following strategies to deal with low power:
- Big effects: We applied meta-experiments only for measures where we expected a big effect based on observed user behavior, user support questions, and user research.
- Longer runtime: Our meta-experiments ran longer than most of our web experiments, a few months as compared to a few weeks. Longer runtimes did not only help increase sample size, but also made sure that most of the experiments in our sample could run to completion.
- Zoom in on the treated population: We only included samples which had a problem and which would receive the treatment if in variant. This reduces the sample size (compared to all experiments, for example), but this is outweighed by the reduction in noise from untreated samples.
- Sensitive metrics: We used binary metrics that measured parts of the KPI that were impacted, rather than the full KPI itself, and in some cases we fell back to behavioral metrics which are easier to move and still indicated an effective process change.
- Better methods: Power can be increased by using covariate adjustment and sequential testing. The latter also helps when there are no clear expectations for treatment effects, because you can afford a design with a smaller minimum detectable effect and stop early if the actual effect is larger.

## Outcome: Metrics choice, and meta-experiment results

The success metrics of an experiment obviously depend on the KPI that should be improved as well as the treatment. We like to choose an operational sub-part of the KPI that is targeted by the intervention, such as a specific user satisfaction score component, or the number of teams that create a draft experiment.

In the meta-experiment of low-power alerts, we defined our success metric as whether an experiment had sufficient power. Since we only included experiments in our meta-experiment which were underpowered at treatment time, this binary metric measured directly our success. We measured this metric once daily for all running experiments included in our meta-experiment.

We also considered measuring power when the experiment ended, or only for shipped experiments. However, this could introduce bias, because our intervention potentially changes the duration and decision of experiments: In variant, experiments could run longer in order to achieve sufficient power, or ship more likely due to being better at detecting true effects. We included metrics on this as monitoring metrics. We also defined supporting metrics for our hypothesis: that experimenters would increase experiment power by clicking the link in our alert, and changing the experiment settings according to our suggestions.

Table 1 shows the outcomes on these metrics for our meta-experiment, corroborating our hypothesis: About 40% of experimenters who received the alert clicked on it, and 15% more experiments changed power-related parameters of their experiments. 10% more experiments had sufficient power. All these effects were significant. Further metrics, such as whether experiments were shipped or not, did not change significantly. We also investigated heterogeneous treatment effects, to see whether our treatment worked better for some experiment types than others, but could not find any significant differences, possibly due to not enough power of our meta-experiments for these smaller subgroups.

| Metric type | Metric | Absolute effect | Significance |
| --- | --- | --- | --- |
| Success metric (KPI component) | Experiment has sufficient power | 10% | Significant (p=0.0045) |
| Supporting behavioral metric | Clicked link in alert (only possible in variant) | 40% | N/A |
| Supporting behavioral metric | Changed power-related experiment settings | 15% | Significant (p=0.0008) |
| Monitoring metric | Experiment not shipped | 0.15% | Not significant (p=0.98) |

*Table 1: Metrics and their results for the meta-experiment of sending alerts to experimenters when their experiments were underpowered at 1 week of runtime*

After the conclusion of the meta-experiment, we reached out to experimenters. We received positive feedback on the intervention. As expected, experimenters who responded to the treatment found the alerts and in particular the one-click setting change helpful. Some started to rely on these alerts to set power on their experiments. Non-responders said that they did not see the alerts, or that they preferred to stick to their experiment settings for reasons unrelated to power, giving us ideas for future interventions.

Due to spillover between variants (some experimenters ran multiple experiments during our meta-experiments), we expect some bias in our quantitative measurements. We believe that this bias "amplified" our effects, since experimenters started to rely on power alerts to set power and therefore might have missed power problems in experiments that were part of base. We therefore think that our quantitative success measure might be an overestimation, but that the decision to ship the meta-experiment is correct.

## Experimenting on colleagues, and dogfooding for experimentation teams

For meta-experiments, the "subjects" of experimentation are our colleagues. Before starting our first meta-experiment, we notified experimenters that we would start running experiments on our in-house experimentation platform. We received positive feedback from a few experimentation enthusiasts, and no negative feedback on this announcement, a sign that experimentation is widely accepted in our company.

We ran our meta-experiments on our own in-house experimentation platform, together with the experiments that were its samples. Note that our meta-experiment about low-power alerts was not included as a sample in itself, because we made sure that our meta-experiment was sufficiently powered. All experiments of our "democratized" experimentation platform are open for all experimenters in the company to view. Since we were worried of biasing experimenters in a "non-blind" study, we hid the display of our meta-experiment behind a feature flag while it was running. We answered user support questions about the low-power alerts without mentioning the meta-experiment. Thus we successfully "hid" the meta-experiment from all experimenters, except for a few experimentation enthusiasts who followed up on our meta-experiment announcement (most of these did not have experiments in the meta-experiment, as they are among our "high-quality" experimenters).

While our experimentation team has deep knowledge about experimentation and our experimentation platform, many members of the experimentation team had not actually run experiments themselves previously. Running a meta-experiment provided a good dogfooding experience by using our own product for a "real" experiment. Having to choose experimentation units and metrics, struggling with low sample size, overall using the UI of our own experimentation platform "for real", and being emotionally invested in the experiment, brought home pain points that we knew from our user support and user research but had not experienced ourselves. The experience increased our empathy for our users, which will help us in user support and future development of our experimentation platform.

## Conclusion

We showed how we improved the experimentation process through "meta-experiments"; by running experiments on experiments or experimenters. In our example meta-experiment of alerting experimenters when their experiments were underpowered, we significantly increased the number of sufficiently powered experiments. Running an A/B test allowed us to causally link our intervention to this increase. This would not have been possible by time series analysis, e.g. using the number of properly powered experiments over time, because other factors like experiment training or the type of experiments run influence experiment power. In addition, while our effect was large enough to be detectable in a proper randomized controlled trial on specifically experiments with problems, the effect was not clearly visible in the timeline due to too many fluctuations with the relatively low number of experiments over time.

Running meta-experiments comes with challenges. Due to relatively small sample sizes when experimenting on experiments or experimenters, long runtimes or large effect sizes are required for sufficient power, which can slow down the velocity of the experimentation team. Another challenge arises from biases and potential user confusion due to variant spillovers when experimenters get exposed to different variants of the experimentation process. We therefore run only a few of our interventions as meta-experiments: when we aim to move a KPI (as opposed to stack modernization or bug fixes, for example) and when we are particularly interested in causal attribution or impact estimation. Additional criteria are that we expect large enough effect sizes to be detectable in a reasonable amount of time, and that the intervention is not confusing to experimenters should they see the treatment of different variants.

While running meta-experiments can be challenging and consumes effort and time, we think that experimenting on the experimentation process is worth it. While the main reasons are causal attribution and impact quantification, we also appreciate the "dogfooding" effect for the experimentation team of running their own experiments, in order to increase product knowledge and user empathy.

## Acknowledgements

Running meta-experiments on our experimentation platform were a team effort. Many thanks to Carolin Grahlmann and Nils Skotara for the meta-experiment design; Alberto de Murga, Chad Davis, Cyprian Mwangi, Joao Arruda, Sebastian Sastoque for the meta-experiment implementation; and Sergey Alimskiy for the meta-experiment UX design, and also the pictures in this article.